# A flexible and accurate total variation and cascaded denoisers-based image reconstruction algorithm for hyperspectrally compressed ultrafast photography

Zihan Guo[1,#], Jiali Yao[1,#], Dalong Qi[1,*], Pengpeng Ding[1], Chengzhi Jin[1], Ning Xu[1], Zhiling Zhang[1], Yunhua Yao[1], Lianzhong Deng[1], Zhiyong Wang[2], Zhenrong Sun[1], Shian Zhang[1,3,4,*]

[1]*State Key Laboratory of Precision Spectroscopy, School of Physics and Electronic Science, East China Normal University, Shanghai 200241, China.*

[2]*School of Mathematical Sciences, University of Electronic Science and Technology of China, Chengdu 611731, China.*

[3]*Collaborative Innovation Center of Extreme Optics, Shanxi University, Taiyuan 030006, China.*

[4]*Joint Research Center of Light Manipulation Science and Photonic Integrated Chip of East China Normal University and Shandong Normal University, East China Normal University, Shanghai 200241, China.*

[#]*The authors contributed equally to this work.*

[*]*Corresponding authors: dlqi@lps.ecnu.edu.cn; sazhang@phy.ecnu.edu.cn*

**Abstract**

Hyperspectrally compressed ultrafast photography (HCUP) based on compressed sensing and the time- and spectrum-to-space mappings can simultaneously realize the temporal and spectral imaging of non-repeatable or difficult-to-repeat transient events passively in a single exposure. It possesses an incredibly high frame rate of tens of trillions of frames per second and a sequence depth of several hundred, and plays a revolutionary role in single-shot ultrafast optical imaging. However, due to the ultra-high data compression ratio induced by the extremely large sequence depth as well as the limited fidelities of traditional reconstruction algorithms over the reconstruction process, HCUP suffers from a poor image reconstruction quality and fails to capture fine structures in complex transient scenes. To overcome these restrictions, we propose a flexible image reconstruction algorithm based on the total variation (TV) and cascaded denoisers (CD) for HCUP, named the TV-CD algorithm. It applies the TV denoising model cascaded with several advanced deep learning-based denoising models in the iterative plug-and-play alternating direction method of multipliers framework, which can preserve the image smoothness while utilizing the deep denoising networks to obtain more priori, and thus solving the common sparsity representation problem in local similarity and motion compensation. Both simulation and experimental results show that the proposed TV-CD algorithm can effectively improve the image reconstruction accuracy and quality of HCUP, and further promote the practical applications of HCUP in capturing high-dimensional complex physical, chemical and biological ultrafast optical scenes.

## 1. Introduction

Single-shot ultrafast optical imaging (UOI) is an advanced research field aiming at visualizing non-repeatable or difficult-to-repeat transient events in a single exposure with imaging speeds of more than billions of frames per second (fps) and multiple frames[1,2]. As one of the cutting-edge passive imaging technologies achieving the incredibly high frame rate of tens of trillions of fps and a sequence depth of several hundred, the compressed sensing-based (CS) UOI technology has developed into a powerful tool in the field of single-shot UOI in the last decade. To date, various CS-UOI techniques have successfully realized the capture of flying photons[3,4], the measurement of ultrafast optical fields[5-7], the ultrafast phase-sensitive imaging of transparent objects[8,9], and the real-time observation of optical chaos[10] and optical rogue waves[11], and so on.

CS-UOI is a combination of compressed sensing and streak imaging, in which a spatially undersampled dynamic scene is firstly compressed into a two-dimensional (2D) image by streak imaging, and then an image reconstruction algorithm is applied to recover the original high-dimensional information from the acquired 2D image. At present, the techniques in this category are developing from temporal imaging to multidimensional imaging. For instance, the compressed ultrafast photography (CUP)[3] and the compressed ultrafast spectral-temporal photography (CUST)[12] can acquire spatiotemporal $x$-$y$-$t$ three-dimensional (3D) intensity information, and the compressed optical field topography (COFT)[13] can obtain spatiotemporal $x$-$y$-$t$ intensity and phase information. The hyperspectrally compressed ultrafast photography (HCUP)[14] and the compressed ultrafast spectral photography (CUSP)[15-16] can acquire spatial-temporal-spectral $x$-$y$-$t$-$\lambda$ four-dimensional (4D) intensity information, and the ultrafast light field tomography (LIFT)[17] can resolve volumetric-temporal $x$-$y$-$z$-$t$ 4D intensity information. Moreover, the volumetric-spectral CUP (SV-CUP) can acquire volumetric-temporal-spectral $x$-$y$-$z$-$t$-$\lambda$ five-dimensional (5D) intensity information[18] and the stereo-polarimetric CUP (SP-CUP) can capture volumetric-temporal-polarimetric $x$-$y$-$z$-$t$-$\varphi$ 5D intensity information[19]. These techniques can be classified into two categories based on the way of streak imaging, one is to form the streak only along a single direction. For example, CUP, SP-CUP and LIFT perform the temporal shearing of a dynamic scene only along the vertical direction by using streak cameras, and CUST and COFT perform the spectral (temporal) shearing of a dynamic scene illuminated by a temporally chirped pulse only along the horizontal direction by using gratings. The other one is to form streaks along two orthogonal directions, including HCUP and CUSP, which perform temporal shearing and spectral dispersion independently and orthogonally. For convenience, we use CUP to represent the former, and HCUP to represent the latter. Obviously, HCUP can maximally access to the information carried by different photon tags benefitting from its two independent shearing directions. Currently, HCUP has been successfully used for the measurement of a chirped picosecond laser pulse[14], spectrally resolved fluorescence lifetime imaging microscopy[15], and laser-induced filament observation[16], etc.

The image reconstruction algorithm is as important as the data acquisition hardware in computational imaging, CS-UOI techniques are no exception. To improve the image reconstruction quality of CUP, a series of advanced algorithms have been developed as the alternatives of the primary two-step iterative shrinkage/thresholding (TwIST)[20] algorithm in recent years. For example, Yang *et al*. developed a hybrid augmented Lagrangian (AL)- and deep learning-based algorithm to optimize the sparse domain and relevant iteration parameters, which apparently improves the efficiency and accuracy of the image reconstruction[21]. Ma *et al*. developed an end-to-end deep learning algorithm capable of recovering the dynamic scene with sharper boundaries, higher feature contrast, and fewer artifacts[22]. Yao *et al*. developed a total variation (TV) combined 3D block-matched filtering algorithm that can simultaneously utilize gradient sparsity and non-local similarity for the image reconstruction, which not only improves the reconstruction quality, but also strengthens the noise immunity of CUP[23]. Jin *et al*. developed a multi-channel coupled and multi-scale weighted denoising algorithm based on the plug-and-play (PnP) framework, which effectively improves the accuracy and quality of reconstructed images by removing non-Gaussian distributed noises using the weighted multi-scale denoising strategy[24]. Compared with the imaging modality of CUP, the increase of an additional shearing direction makes the data compression ratio of HCUP show about an order of magnitude increasement, which leads to the fact that the two mainstream iterative algorithms currently used in HCUP, i.e., the TwIST and AL[25] algorithms, can no longer meet the reconstruction requirements. These two algorithms are based on the TV regularization. However, this regularization only takes into account the local features of the images, which is insufficient to express the priori information. It is for this reason that TV-regularized algorithms usually produce the staircasing phenomenon in the smoothing region and result in the loss of the image details[26]. Therefore, the quality of the reconstructed image deteriorates rapidly in the face of the increase of the data compression ratio. Unfortunately, no advanced image reconstruction algorithm has been applied to HCUP so far.

To solve this problem, we propose a new image reconstruction algorithm based on TV and cascaded denoisers (CD) for HCUP, referred to as the TV-CD algorithm. The algorithm adapts a PnP-based alternating direction method of multipliers (PnP-ADMM) as an iterative framework and jointly uses the TV denoising model along with three advanced deep learning-based denoising models, including the fast and flexible denoising convolutional neural network (FFDNet), the dilated-residual U-Net denoising neural network (DRUNet) and the fast deep video denoising network (FastDVDNet) in the framework, which can preserve the image smoothness while utilizing the deep denoising networks to obtain more priori and solve the common sparsity representation problem in local similarity and motion compensation. By comparing the TV-CD algorithm with the traditional TwIST and AL algorithms in both numerical simulation and experimental measurement results, the superiority of the proposed algorithm in HCUP image reconstruction is proved. It can be prospected that TV-CD can vastly promote the practical applications of HCUP in capturing high-dimensional information of ultrafast optical scenes.

## 2. Basic Principles

As mentioned above, HCUP can be divided into the data acquisition and image reconstruction. The forward data acquisition process is shown in Fig. 1(a). Firstly, a 4D scene $I(x, y, t, \lambda)$ with spatial-temporal-spectral information is spatially undersampled by a pseudo-randomly distributed binary matrix, which is typically a mask generated by a spatial light modulator. Subsequently, different spectral components of the encoded scene are dispersed by a prism or grating, and different temporal frames of the coded scene are deflected orthogonally to the direction of spectral dispersion, typically by an ultrafast streak camera or electro-optical deflector[27]. Finally, the encoded, dispersed and deflected 4D scene is integrated and recorded as a 2D image $E(x', y')$ through the spatial-temporal-spectral integration via a planar array detector. Overall, this data acquisition process can be mathematically formulated as[14]

$$E(x', y') = \mathbf{MTSC}I(x, y, t, \lambda), \tag{1}$$

where **C** is the spatial encoding operator, **S** is the spectral shearing operator, **T** is the temporal shearing operator, and **M** is the spatial-temporal-spectral integration operator.

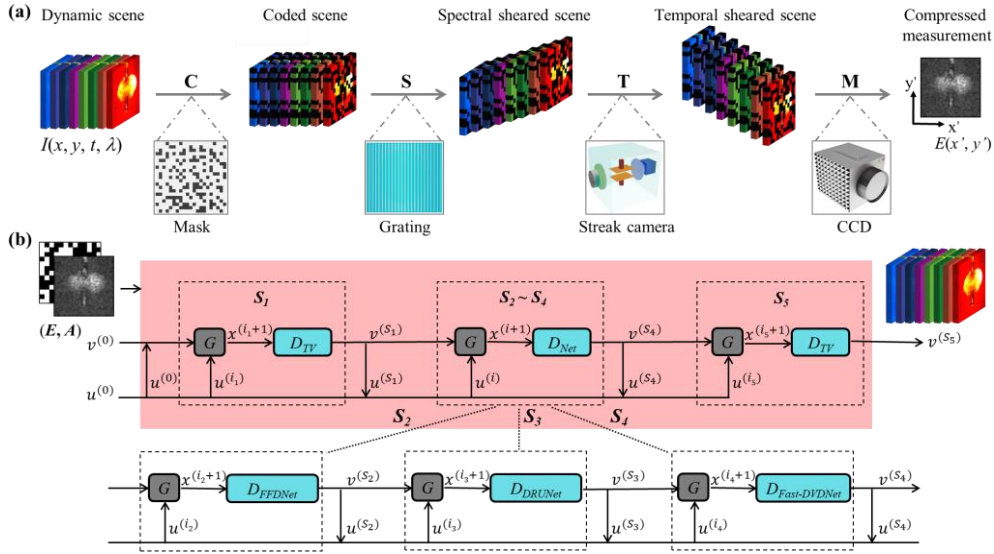

**Figure 1** (a) Schematic diagram of the data acquisition for HCUP, where $t$: time; $\lambda$: spectrum; $x$, $y$: spatial coordinates of the dynamical scene; $x'$, $y'$: spatial coordinates at the detector. (b) Data flowchart of the TV-CD algorithm for HCUP reconstruction, where a 5-step cascaded denoising of [TV, FFDNet, DRUNet, FastDVDNet, TV] is used and each step is composed of the projections $G$ and $D$ representing the operations in Eqs. (8) and (9), respectively.

In the inverse image reconstruction process, the original scene is recovered based on the observed 2D image $E(x', y')$ and the shifted 4D sampling cube, i.e., solving the inverse problem of Eq. (1). However, the inverse problem is ill-conditioned, and traditional optimization algorithms based on compressed sensing usually use a regularization term Φ(*I*) to confine the solution of this inverse problem to the desired signal space, where the optimal estimation $I^*$ of *I* can be found by minimizing an objective function *f*(*I*), as follows[28-30]:

$$I^* = \arg\min_I f(I) = \arg\min_I \frac{1}{2}\|E - \mathbf{MTSC}I\|_2^2 + b\Phi(I), \qquad (2)$$

where $\|\cdot\|_2$ is the $l_2$ norm, $\|E - \mathbf{MTSC}I\|_2^2$ is the fidelity term presenting that the reconstructed scene needs to conform to the sampling equation, Φ(*I*) is the regularization term representing that the reconstructed scene needs to satisfy the prior, and *b* is the regularization parameter to balance the two terms. For convenience, **A** = **MTSC**. Here, we propose a TV-CD algorithm based on the PnP-ADMM framework to solve problem (2). In the PnP framework[31,32], by introducing an auxiliary variable *v*, problem (2) can be rewritten as

$$\arg\min_{I,v} \left\{\frac{1}{2}\|E - \mathbf{A}I\|_2^2 + b\Phi(v)\right\} \quad s.t. \ v = I. \qquad (3)$$

According to the AL multiplier method, the optimal solution of problem (3) can be obtained by minimizing an AL function, expressed as follows:

$$\arg\min_{I,v,u} \frac{1}{2}\|E - \mathbf{A}I\|_2^2 + b\Phi(v) + u^T(v - I) + \frac{\rho}{2}\|v - I\|_2^2, \qquad (4)$$

where *u* is another auxiliary variable, and *ρ* is the penalty parameter. Furthermore, problem (4) can be solved by the ADMM algorithm, which can be decomposed into the following sequence of subproblems[33]:

$$I^{k+1} = \arg\min_I \frac{1}{2}\|E - \mathbf{A}I\|_2^2 + \frac{\rho}{2}\left\|I - \left(v^k - \frac{1}{\rho}u^k\right)\right\|_2^2, \qquad (5)$$

$$v^{k+1} = \arg\min_v b\Phi(v) + \frac{\rho}{2}\left\|v - \left(I^k + \frac{1}{\rho}u^k\right)\right\|_2^2, \qquad (6)$$

$$u^{k+1} = u^k + \rho\left(I^{k+1} - v^{k+1}\right). \tag{7}$$

Here, the superscript *k* represents the *k*-th iteration. Subproblem (5) is an equation of quadratic form that has a closed-form solution shown as[34]

$$I^{k+1} = \left[\mathbf{A}^T\mathbf{A} + \rho I\right]^{-1}\left[\mathbf{A}^T E + v^k - \frac{u^k}{\rho}\right], \tag{8}$$

and subproblem (6) is a denoising problem based on the Φ(*v*) regularization, where $\left(I^k + \frac{1}{\rho}u^k\right)$ is a noisy image and *v* is a denoised image. The problem can be solved by using off-the-shelf denoising algorithms, and can be formulated as[32]

$$v^{k+1} = D_\sigma\left(I^k + \frac{1}{\rho}u^k\right), \tag{9}$$

where *D* is the denoising algorithm being used and $\sigma$ is the noise standard deviation. As the TwIST and AL algorithms commonly applied to HCUP so far are both based on the TV regularization, the TV denoising algorithm is also considered for this subproblem. However, TV denoisers based on local smoothness prior tend to introduce staircasing artifacts in the reconstruction, which greatly limits the image reconstruction quality. Inspired by recent advances in deep learning-based denoisers[7,24] and combinations of multiple conventional denoisers[23,35] within the PnP framework to achieve image quality improvement in CUP, we combine the traditional TV denoiser with three advanced deep learning-based denoisers in series to solve subproblem (6), including FFDNet[36], DRUNet[37] and FastDVDNet[38]. Here, FFDNet and DRUNet can remove the noise in a single frame out of the dynamic scene by taking the noise level maps as inputs. Complementarily, FastDVDNet effectively utilizes the information in the temporal neighborhood and strengthens the temporal correlation of the remaining noise in the output frames to further improve the overall denoising effect of the dynamic scene. Theoretically, different denoisers represent the utilization of different priors, and these deep learning networks driven by a large amount of data can acquire richer priors to better recover the intrinsic features of the structures in the image or scene. Fig. 1(b) shows the flowchart of the image reconstruction for HCUP via the TV-CD algorithm. Here, the 2D observed image $E(x', y')$ and the combined linear operator **A** are input into the PnP-ADMM iterative framework, where the projection $G$ represents the operation in Eq. (8), and the projection $D$ represents the operation in Eq. (9). We use a cascaded denoising configuration with [TV, FFDNet, DRUNet, FastDVDNet, TV] and assign a separate number of iterations to each denoiser. When the number of iterations of one denoiser is completed, other denoisers are replaced in the subsequent iterations, and the reconstructed 4D scene is output until all of the iterations are finished. In each iteration, the noise standard deviation $\sigma$ is adaptively updated by considering the relative residue and monotonously decreased[39].

## 3. Theoretical Simulations

In order to validate the performance of the TV-CD algorithm, three different types of scenes were compressed and reconstructed for simulation. In the first two types of scenarios, several multispectral and video scenes were selected, in which the images differ from each other in only one dimension. Moreover, the images in the third type of scenario are variable in both the temporal and spectral dimensions. According to the HCUP image acquisition process, all scenes were first encoded with a {0, 1} pseudo-random spatial encoding for each image, with the pixel size of the mask being the same as that of the detector. Then, the encoded images with different spectra in each scene were sequentially shifted along the lateral direction ($x$-direction), with each image shifted to the right by one pixel relative to the previous one to simulate the spectral dispersion. Similarly, the images at different time instants were sequentially shifted along the vertical direction ($y$-direction), and each frame was shifted downward by one pixel with respect to the previous one to simulate the temporal deflection. Finally, all the images were superimposed to obtain a 2D observation image. In each simulation, the proposed TV-CD algorithm, as well as the mainstream TwIST and AL algorithms were employed to recover the original 4D scenes from the 2D observed images, respectively. For each algorithm, the average peak signal-to-noise ratio (PSNR) and structural similarity (SSIM) values of all the reconstructed images were calculated based on the ground truths as quantitative image quality assessments (IQAs).

**Table 1** The averaged results of PSNR (dB) and SSIM by different algorithms on multispectral and video scenes.

| Data | IQA | Methods | | | Data | IQA | Methods | | |
|---|---|---|---|---|---|---|---|---|---|
| Multispectral Scenes | | TwIST | AL | TV-CD | Video Scenes | | TwIST | AL | TV-CD |
| Clay | $\overline{PSNR}$ | 20.64 | 21.00 | 27.73 | Welding | $\overline{PSNR}$ | 18.81 | 19.39 | 22.46 |
| | $\overline{SSIM}$ | 0.7011 | 0.7704 | 0.9141 | | $\overline{SSIM}$ | 0.5733 | 0.6830 | 0.8174 |
| Flowers | $\overline{PSNR}$ | 26.18 | 25.76 | 27.32 | Detonators | $\overline{PSNR}$ | 18.94 | 19.71 | 24.19 |
| | $\overline{SSIM}$ | 0.6281 | 0.6701 | 0.8255 | | $\overline{SSIM}$ | 0.6663 | 0.7572 | 0.8788 |
| Painting | $\overline{PSNR}$ | 25.44 | 24.01 | 28.28 | Snowboard | $\overline{PSNR}$ | 20.95 | 21.47 | 24.08 |
| | $\overline{SSIM}$ | 0.5416 | 0.6077 | 0.8198 | | $\overline{SSIM}$ | 0.7372 | 0.7792 | 0.8601 |
| Pompoms | $\overline{PSNR}$ | 24.47 | 25.35 | 29.53 | Tennis | $\overline{PSNR}$ | 20.36 | 19.16 | 23.76 |
| | $\overline{SSIM}$ | 0.6878 | 0.7566 | 0.9156 | | $\overline{SSIM}$ | 0.5915 | 0.6631 | 0.8008 |
| Spools | $\overline{PSNR}$ | 23.48 | 24.45 | 26.50 | Filament | $\overline{PSNR}$ | 23.10 | 22.80 | 24.26 |
| | $\overline{SSIM}$ | 0.6856 | 0.7293 | 0.8833 | | $\overline{SSIM}$ | 0.8470 | 0.8985 | 0.9360 |
| Average | $\overline{PSNR}$ | 24.04 | 24.11 | **27.87** | Average | $\overline{PSNR}$ | 20.43 | 20.51 | **23.75** |
| | $\overline{SSIM}$ | 0.6488 | 0.7068 | **0.8717** | | $\overline{SSIM}$ | 0.6831 | 0.7562 | **0.8586** |

Firstly, five multispectral scenes including *Clay*, *Flowers*, *Painting*, *Pompoms* and *Spools* from the CAVE dataset[40] were selected to examine the performance of the TV-CD algorithm for spectral scenes, and each of them contains a data size of 256×256×8×8 with 8 temporal and 8 spectral channels. It is worth noting that each scene varies in the spectral dimension but keeps unchanged in the temporal dimension. Each scene was forward compressed based on the image acquisition settings and then inversely reconstructed with the TwIST, AL and TV-CD algorithms, respectively. The IQA values of the reconstruction results by the three algorithms are displayed on the left side of Table 1, including the PSNR and SSIM. As can be seen, the reconstruction result by the TV-CD algorithm yields much higher IQA values than those of the TwIST and AL algorithms for each scene. In addition, the IQA values for all the five scenes are averaged to obtain average PSNR (SSIM) improvements of 3.83 dB (0.223) and 3.76 dB (0.165) for the TV-CD algorithm over the TwIST and AL algorithms, respectively, which are shown in bold font. For a more visual comparison, the ground truth datacube and reconstructed results of *Flowers* are shown in Fig. 2(a). For simplicity, only the second, fifth and eighth frames in both the temporal and spectral channels are displayed, and the values below each sub-image are the corresponding PSNR (dB) and SSIM. In addition, specific regions are selected for zoomed-in display. It can be obviously seen that although the TwIST algorithm is able to recover the spectral evolution of the scene, the reconstructed images are too blurred with serious loss of detailed information. What's worse, the AL algorithm fails to recover the spectral evolution information. In contrast, the TV-CD algorithm not only correctly reconstruct the spectral evolution, but also obtain clearer structural boundaries and more image details.

Similarly, five video scenes including *Welding*, *Detonators*, *Snowboard*, *Tennis* and *Filament* captured by ultrahigh-speed cameras were also selected to examine the performance of the TV-CD algorithm for dynamic scenes, in which the data size of each scene is the same to that of a multispectral scene. Nevertheless, each scene varies in the temporal dimension but keeps unchanged in the spectral dimension. The average PSNR and SSIM values of the five video scenes reconstructed by the three algorithms are also listed on the right side of Table 1, in which the TV-CD algorithm still shows superiority over the other two algorithms. Moreover, the averaged PSNR (SSIM) value of the TV-CD algorithm for all the video scenes shown in bold font is 3.32 dB (0.175) and 3.24 dB (0.102) higher than those of the TwIST and AL algorithms, respectively. As a visual comparison, Fig. 2(b) shows representative frames of the ground truth and reconstruction results of the *Detonators* with the same temporal and spectral indices as in Fig. 2(a). It can still be observed that the reconstructed images of both the TwIST and AL algorithms show serious artifacts, which can hardly recover the actual dynamics of the gradually expanding shockwave generated by the detonator explosion. On the contrary, the TV-CD algorithm can recover the characteristics of the shock wave propagation in rather high fidelity, which can be clearly distinguished by the information in the magnified box.

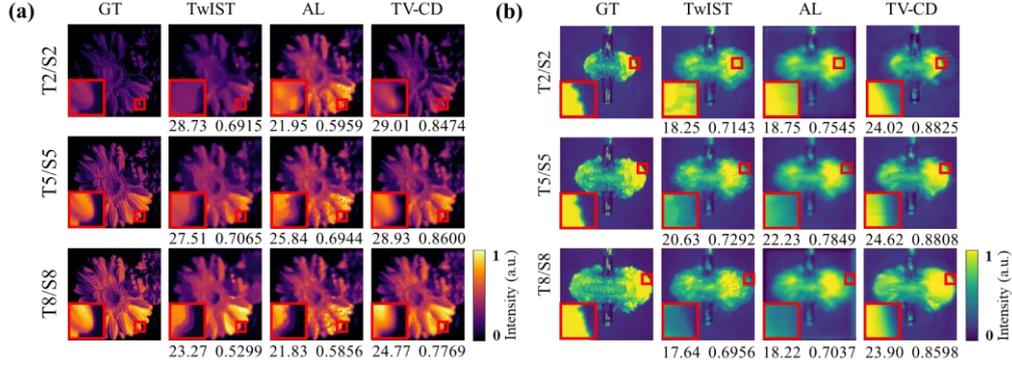

**Figure 2** (a) and (b) Representative reconstructed frames of *Flowers* and *Detonators* by the algorithms, respectively, together with the ground truths (GT) for comparison. The sub-image in the bottom left corner of each frame is the enlarged scene in the corresponding red box.

As a reconstruction algorithm for HCUP, the performance of dealing with simultaneously temporal and spectral varying scenes is very important. Therefore, a 4D spectral-temporal-varying scene containing 10 temporal and 10 spectral channels with each frame size of 358×358 was further generated for simulation. The scene consists of a triangle, a circle and a rectangle positioned with fixed relative distances between each other. However, the shapes not only show different intensities in different spectral channels, but also rotate globally with a constant speed within the temporal frames. Ten representative frames out of the overall 100 frames of the scene with the same indices in the temporal and spectral dimensions are shown in Fig. 3(a). As can be seen, these three shapes in each image are rotated counterclockwisely by 18° with respect to the previous one in the temporal dimension. In the spectral dimension, the intensities of the shapes evolve according to a self-defined law independently. Fig. 3(b) shows the intensity evolution settings for each shape in different spectral channels, where the intensity within any shape is uniform. With the intensity normalization in consideration, 1 denotes the normalized maximum intensity and 0.1 denotes an intensity that is one-tenth of the maximum intensity. In the spectral channels, the intensity of the triangle monotonously weakens from 1 to 0.1 with a fixed intensity difference of 0.1. For the circle, the intensity gets stronger and then weaker, where the strongest value is only half of that of the triangle at 0.5. For the rectangle, its intensity evolution is just the opposite of that of the triangle. Representative five frames (indicated by T$i$/S$i$ ($i$=2, 4, 6, 8, 10)) of the reconstructed results obtained by the TwIST, AL and TV-CD algorithms are shown in Fig. 3(c), respectively. It can be seen that the TwIST and AL algorithms can only roughly recover the shape features with high intensity, such as the triangle in T2/S2 and the rectangle in T10/S10. However, they fail to recover the shape features with low intensity, such as the circle and rectangle in T2/S2, and the circle and triangle in T10/S10. In contrast, the TV-CD algorithm not only clearly recovers all the shape features of the scene, but also correctly reconstructs the spectral-temporal intensity evolution of the scene. For quantitative comparison, the average PSNR and SSIM values of all the reconstructed frames by the three algorithms are further calculated, and the results are shown in Fig. 3(d). As can be seen, the results by the TV-CD algorithm has an average PSNR value of up to 34.98 dB, far superior to the other two, and an average SSIM value of 0.867, which is also the best value among the three algorithms. In addition, the centroid X-Y coordinates of the shapes with distinguishable contours in the reconstruction results were also extracted for comparison, and the results are shown in Figs. 3(e) and (f), in which the ten representative frames are indexed sequentially. It can be clearly seen that the temporal evolution of the centroid coordinates of the three shapes reconstructed by the TV-CD algorithm is the most consistent with those of the ground truth, while the TwIST and AL algorithms have anomalous coordinate points that significantly deviate from the ground truth. It is worth noting that restricted by the abilities of dealing with scenes with low SNR of the TwIST and AL algorithms, several shapes are failed to reconstruct, such as the circle in T1/S1 and T2/S2 reconstructed by the AL algorithm, as well as the triangle in T8/S8 and T9/S9 reconstructed by the TwIST algorithm. The results suggest an excellent performance in terms of the image quality and measurement accuracy of the TV-CD algorithm.

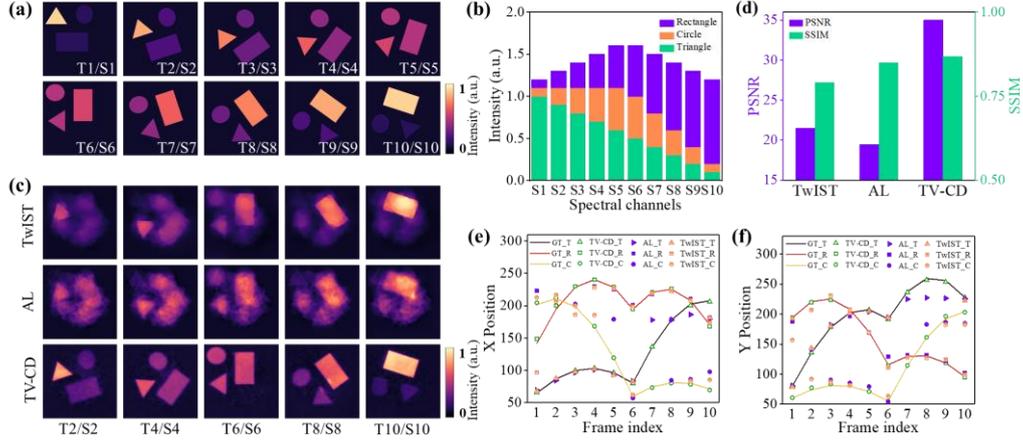

**Figure 3** (a) Ten representative ground truth frames of the 4D spatial-temporal-spectral scene. (b) Intensity settings of each shape in the spectral dimension. (c) Representative reconstructed frames of the algorithms. (d) Average PSNR and SSIM values of the reconstructed results. (e) and (f) Comparison of the centroid positions of the three shapes in reconstruction to the ground truth in the X and Y directions, respectively. T: triangle, R: rectangle, C: circle.

## 4. Experimental Results

Furthermore, the TV-CD algorithm was employed to reconstruct the experimentally obtained ultrafast laser field and photo-induced fluorescence dynamics scenes by a home built HCUP system, and further visualize the simultaneously spatial-temporal-spectral resolved ultrafast scenes. As reported previously, the temporal and spectral frame intervals of the HCUP system are 2 ps and 1.72 nm, respectively[14]. In experiment, a temporally chirped picosecond laser pulse was firstly measured by the HCUP system, and the configuration of the experiment is shown in Fig. 4(a). A mode-locked Ti:Sapphire laser amplifier is used to generate femtosecond laser pulses with a center wavelength of 800 nm. An output laser pulse is first frequency chirped by a pulse stretching device (PSD), then an E-shaped mask is used to spatially modulate the chirped pulse, and the chirped E-shape pulse is then compressed and acquired by the HCUP system. In the HCUP system, the chirped E-shaped laser pulse firstly passes through a camera lens (CL) and a lens L1 to form an intermediate scene in the plane of a transmissive spatial encoder with a pseudo-random binary pattern, and the spatially encoded scene is then transmitted by an optical 4f system consisting of lenses L2 and L3 to a streak camera with a fully opened entrance slit. A transmissive grating (G) is placed in front of the streak camera to disperse the scene in spectrum horizontally. In the streak camera, photons in the laser pulse are first converted to electrons by a photocathode, and the accelerated electron pulse is subsequently operated by a vertical sweeping voltage for temporal deflection. Finally, the electrons are multiplied by a microchannel plate and bombard on a phosphor screen to be converted back into photons, and a 2D measurement image is captured by the internal CCD.

After being compressed and acquired by the HCUP system with a suitable time window of the streak camera, the spatial-temporal-spectral 4D information of this temporally chirped E-shaped laser pulse was reconstructed by using the TwIST, AL and TV-CD algorithms, respectively. Here, the reconstruction yielded a data cube size of 54×54×23×23 for $I(x, y, t, \lambda)$ with a data compression ratio of up to 529:1, and representative frames of the reconstruction results are selected for display in Fig. 4(b). Considering the pulse duration and central wavelength of the pulse, the frames at the time instants of 60, 80, 100, 120 and 140 ps and the spectral channels of 794.18, 797.62 and 801.06 nm are selected. It can be seen that all the three algorithms recover the positive frequency chirp characteristics of the pulse, with the longer wavelength components appearing earlier compared to the shorter ones. However, at such a high data compression ratio, the TwIST and AL algorithms solely based on the TV priori are unable to recover the complete E-shaped spatial structure at frames with low intensities. What's worse, the spatial intensity distribution of the recovered structure is extremely heterogeneous, e.g., in the reconstructed image at 80 ps/801.06 nm by the TwIST and AL algorithms. In contrast, the TV-CD algorithm obtains better reconstruction effect in terms of spatial structure and intensity distribution.

To quantitatively evaluate the image reconstruction quality, a blind/referenceless image spatial quality evaluator (BRISQUE)[41] was introduced. The BRISQUE does not compute distortion-specific features but instead uses scene statistics of mean subtracted contrast normalized coefficients to quantify the loss of image "naturalness" in the image due to the presence of distortions, leading to an overall measurement of quality. Meanwhile, considering the repeatability of the laser pulse, the spectral-temporal unsheared view acquired by the streak camera in static mode was used as the reference image, and all the reconstructed spectral-temporal resolved frames were integrated to calculate the root-mean-square error (RMSE) between the integrated image and the reference image. The values of the BRISQUE and RMSE are shown in Fig. 4(c), and it should be noted that smaller values represent better image qualities in both the indicators. It can be obviously seen that the results reconstructed by the TV-CD algorithm show the smallest BRISQUE and RMSE values of 43.49 and 0.264, which are reduced by 11.43% and 27.41% to those by the TwIST algorithm, as well as 5.49% and 17.49% to those by the AL algorithm, respectively. Moreover, the normalized spatial intensity distributions along the white dashed lines in the reconstructed images at 80 ps/801.06 nm by the three algorithms in Fig. 4(b) were also extracted and compared with that of the reference image, and the results are shown in Fig. 4(d). Obviously, the intensity distribution reconstructed by the TV-CD algorithm is the closest to the reference, while the other two algorithms result in significant intensity degradations in the second and third peaks of the E-shaped structure with low SNR. In addition, in order to further verify the reconstruction accuracy of the three algorithms in both the temporal and spectral dimensions, their intensity evolution curves from all the reconstruction results were extracted for comparison, and the results are shown in Figs. 4(e) and (f), respectively. Here, the temporal intensity evolution measured by the streak camera in one-dimensional (1D) mode[42] and the spectral intensity evolution obtained with a spectrometer are used as reference curves. It can be found that the reconstruction results of the three algorithms are in good agreement with the reference curve for the temporal intensity evolution, achieving a consistent pulse duration in full width at half-maximum (FWHM) of about 100 ps. For the spectral intensity evolution, the reconstruction result of the TV-CD algorithm shows a high fidelity to the reference, achieving a spectral width in FWHM of 16 nm, while the results of the other two algorithms show varying degrees of expansion and deviation.

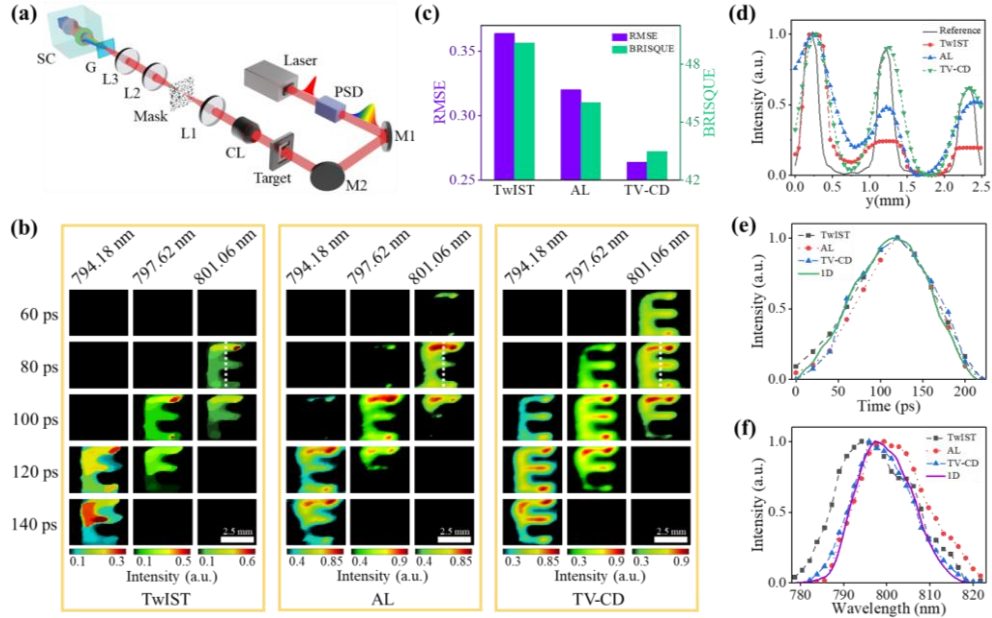

**Figure 4** (a) The experimental configuration for measuring a temporally chirped picosecond laser pulse using an HCUP system. (b) Representative frames of the reconstruction results by the algorithms. (c) The BRISQUE and RMSE values of the reconstruction results by the algorithms. (d) Normalized intensities along the white dotted lines in the reconstructed images at 80 ps/801.06 nm in (b) and the reference. (e) and (f) Extracted temporal and spectral intensity evolutions from the reconstruction results, respectively, together with the reference curves.

Fluorescence lifetime imaging (FLI) is playing an irreplaceable role in photobiological and biomedical applications, and HCUP have shown its promising capability of realizing spectral-resolved FLI in a single shot in recent works[15,18]. Therefore, the second ultrafast scene in experiment is the photo-induced FLI observation of a mannequin model coated with CdSe quantum dots. As shown in Fig. 5(a), a femtosecond laser pulse with a central wavelength of 800 nm and pulse width of 50 fs is output from the Ti:Sapphire laser amplifier for excitation. The central wavelength is converted to 400 nm through the frequency doubling of a barium borate (BBO) crystal, which is within the optical absorption wavelength range of CdSe quantum dots[43, 44]. Subsequently, the pulse is diffused by an engineered scatterer (ED) and reflected by a 425 nm long-pass dichroic mirror (DM) to illuminate a mannequin model coated with CdSe quantum dots. The emitted fluorescence scene then passes through a bandpass filter (BF) and is split into two paths by a beam splitter (BS). The reflected scene is directly imaged by an external CCD camera to form a spectral-temporal unsheared integral reference image, and the transmitted scene is captured by the HCUP system. Similarly, the TwIST, AL and TV-CD algorithms were respectively used to reconstruct the spatial-temporal-spectral 4D information of the fluorescence dynamics for comparison. In addition, the mannequin structure captured by the external camera was used as a spatial constraint to limit the boundary of the reconstructed images. Here, the reconstructed data cube size is 67×105×42×33 for $I(x, y, t, \lambda)$ with the data compression ratio as high as 1386:1. For simplicity, the frames at the time instants of 9.6, 12.0, 14.4, 19.2 and 48.0 ns and the spectral channels of 526.84, 532.00 and 537.16 nm are selected for display in Fig. 5(b). It can be seen that the fluorescence spectra reconstructed by the three algorithms have the same central wavelength of about 532.00 nm in spectrum. In the temporal dimension, the fluorescence intensity rises to a peak at the time instant of 12.0 ns, and then gradually decreases. It is not until 48 ns that the fluorescence intensity reconstructed by the TV-CD algorithm disappears, while the TwIST and AL algorithms still have significant fluorescence intensity. As a quantitative comparison, the BRISQUE and RMSE values of the reconstruction results compared to the reference image recorded by the external camera were calculated, and the results are shown in Fig. 5(c). As expected, the TV-CD algorithm brings the smallest BRISQUE and RMSE values of 53.6736 and 0.2043, which are reduced by 21.29% and 30.86% to those of the TwIST algorithm, as well as 7.68% and 22.02% to those of the AL algorithm, respectively, indicating the best reconstruction quality. The intensity evolution curves in the temporal and spectral dimensions were also extracted from the reconstruction results and compared with the measurements from the streak camera in 1D mode and the spectrometer to confirm the reconstruction accuracy of the three algorithms, and the results of the comparisons are shown in Figs. 5(d) and 5(e), respectively. It is clear that both the temporal and spectral intensity evolutions of fluorescence reconstructed by the TV-CD algorithm are in the best agreement with the reference curve, while the temporal and spectral intensity evolutions by the TwIST and AL algorithms are significantly noisy and broadened. In addition, we also counted the fluorescence lifetimes in different spectral channels, and the lifetimes of some representative spectral components are shown in Fig. 5(f). It can be seen that the lifetimes of different fluorescence spectral components are similar, which suggests that all the fluorescence spectral components come from the relaxation of the same excited state in CdSe quantum dots. Compared to the average fluorescence lifetime of 19.28 ns reconstructed by the TwIST algorithm and 12.61 ns reconstructed by the AL algorithm, the average lifetime reconstructed by the TV-CD algorithm is 8.68 ns, which is in excellent consistent with the lifetime of 8.66 ns measured by the streak camera in 1D mode. It is worth noting that the fluctuations between the lifetimes of different spectral components are minimized in the TV-CD algorithm, demonstrating that this algorithm has a superior reconstruction accuracy.

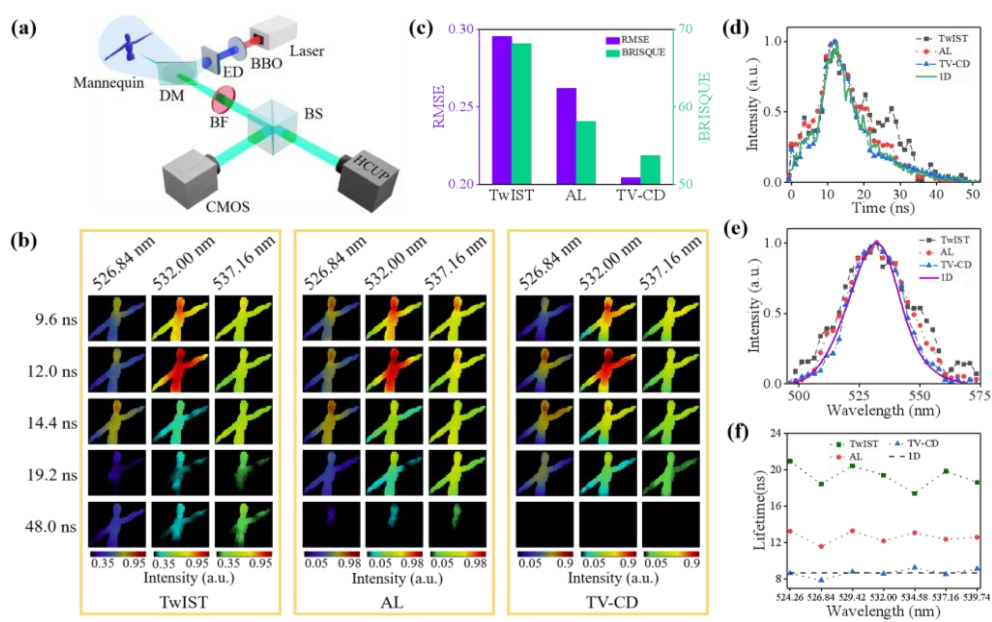

**Figure 5** (a) Experimental configuration for observing a fluorescence dynamical scene using the HCUP system. (b) Representative frames of the reconstruction results by the algorithms. (c) The BRISQUE and RMSE values of the reconstruction results by the algorithms. (d) and (e) Extracted temporal and spectral intensity evolutions from the reconstruction results, respectively, together with the reference curves. (f) Extracted fluorescence lifetimes of representative spectral components of the algorithms, together with the measured result by the streak camera in 1D mode.

## 5. Discussions

Considering the single-shot imaging capability of HCUP-class techniques, as well as its unparalleled advantages in imaging speed, imaging dimension, and sequence depth, the recovery of high-quality scenes from compressed images will undoubtedly continue to gain widespread attention in the future. With the improvement of computing power and the development of new computational frameworks, the emergence of various iterative optimization denoising algorithms and end-to-end deep denoising networks have brought higher reconstruction accuracy and speed to PnP-based algorithms.

First, based on the flexibility of the PnP framework, the type of denoisers can be selected according to the dimensional characteristics of the scene. For example, spectral denoisers[45,46] can be selected for the spectral dimension, and video denoisers[47] can be selected for the temporal dimension. Exploring the optimal combination of multiple denoisers is also one of the critical problems to concern. For instance, in addition to cascaded denoising, a parallel denoising strategy can also be selected in each iteration, and different denoising weights can be set for multiple denoisers to jointly improve the image reconstruction quality of HCUP[48].

Second, further enhancement of the robustness of the PnP framework is necessary. At present, pre-trained denoising networks are plugged into the PnP framework as the priori, and the image reconstruction quality is degenerated when the experimental task differs significantly from the training data of the networks. In perspective, an adaptive PnP framework can be further developed to automatically update parameters in the deep denoising network according to specific dynamic scenes and imaging models to solve the problem of mismatch between pre-trained networks and specific desired scenes[49]. In addition, the combination of the PnP framework and new neural network models will also effectively relieve the pressure of information reconstruction caused by extremely high data compression ratio for HCUP. For example, the new transformer model relies on a self-attention mechanism to avoid the defect of limited receptive field in convolutional neural networks, and can utilize remote long-range non-local similarities to significantly improve the quality of image reconstruction[50,51].

Besides, compared with other imaging techniques based on compressed sensing, such as the coded aperture compressive temporal imaging[52] with much slower imaging speeds and the coded aperture compressive spectral imaging[53] mapping spectrum only, HCUP suffers from the low light throughput resulting in signal drowning in noise when recording ultrafast dynamic scenes with very short durations in a passive imaging mode. Therefore, it is of great significance to develop a deep denoising network for scenes in an extremely low light environment[54]. In addition, advanced deep denoising networks for other photon tags such as spectrum, polarization and phase are also indispensable.

## 6. Conclusion

In summary, we have developed a flexible and accurate image reconstruction algorithm based on TV and cascaded denoisers for HCUP, named the TV-CD algorithm. The algorithm combines the TV and three advanced deep denoising models in an iterative framework of PnP-ADMM, which can robustly recover the inherent features of the structures in ultrafast scenes by utilizing a deep learning network driven by a large amount of data to obtain a richer prior while preserving the smoothness of the image. Through the reconstructions of various types of simulation and experimental scenes, it is proved that the TV-CD algorithm can effectively improve the quality and accuracy of the reconstruction results, including reducing the spatial artifacts and improving the imaging details, compared to the widely used TwIST and AL algorithms. In addition, it is foreseen that the continuous development of the more advanced PnP framework and deep denoising networks will enable PnP-based algorithms to further improve the image reconstruction quality of HCUP in future, which will greatly promote its applications in exploring complex high-dimensional ultrafast scenes.